# Proposal of atomic clock in motion: Time in moving clock

Masanori Sato

*Honda Electronics Co., Ltd.,*
*20 Oyamazuka, Oiwa-cho, Toyohashi, Aichi 441-3193, Japan*

E-mail: msato@honda-el.co.jp

**Abstract**:   The time in an atomic clock in motion is discussed using the analogy of a sing around sound source. Sing around frequency is modified according to the motion of the sing around sound source, using the Lorentz transformation equation. Thus, if we use the sing around frequency as a reference, we can define the reference "time". We propose that the time delay of an atomic clock in motion be derived using the sing around method. In this letter, we show that time is defined by a combination of light speed and motion.

1. INTRODUCTION

   The derivation of the Lorentz transformation equation was clearly described by Feynman et al. [1]. The Doppler shift equation was observed to be different between acoustic wave and light, thus we determined the reason for this difference [2]. We pointed out that the frequency of a sound source should be modified according to its motion. We proposed a sing around sound source whose frequency changes with its velocity, as is suggested by the Lorentz transformation equation.

   We discussed the reference frequency of a moving sound source with respect to the Lorentz transformation equation. The sing around sound source moving in air exhibits a decrease in frequency. If the modified frequency is used as a reference frequency, the time delay in a moving frame can be explained [2].

   We also proposed a number of feasible experiments [3, 4]. These experiments are concerned with signaling by quantum entanglement [3] and interference [4], and may provide for another interpretation of special relativity.

   We clearly accept the theory of special relativity. However, at this stage, we are not satisfied with its current interpretation, because investigating phenomena using its current interpretation is difficult. We believe that special relativity can be defined in a simpler manner. In this study, to define the time delay in motion, we propose a use of the analogy of a sing around sound source in motion.

2. ATOMIC CLOCK IN MOTION

A. Analogy of sing around sound source in motion

In acoustic waves, for the measurement of sound speed, not only interference but also the sing around method is used. The sing around method uses two pairs of transmitters and receivers as shown in **Fig. 1**, where a pulsed signal is transmitted by transmitter 1 and detected by detector 2. After its detection, the pulsed signal is amplified by amplifier 2, and a new pulsed signal is transmitted by transmitter 2, detected by detector 1, transmitted again by transmitter 1, and so on. The frequency of the sing around sound source was determined on the basis of the repetition of pulsed acoustic waves, and detected using the frequency counter.

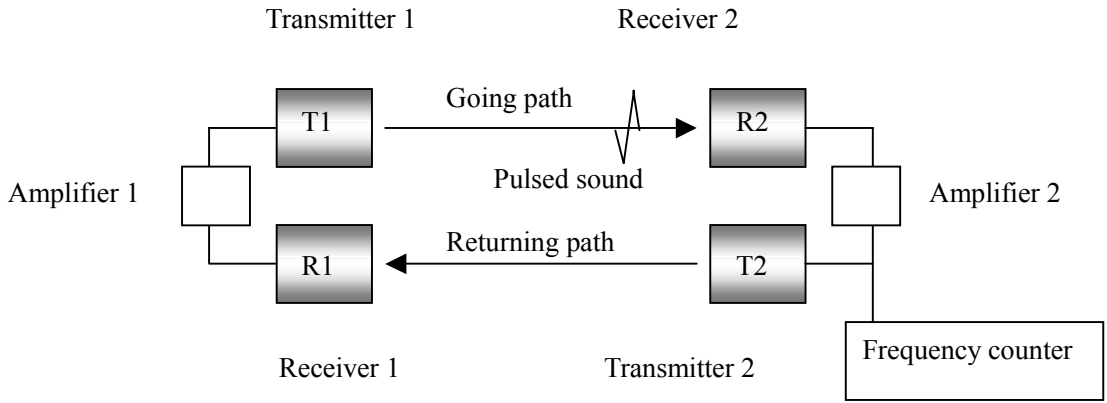

Fig. 1   Sing around sound source at rest.

Now let us consider the frequency modification according to the motion of the sing around sound source. Here, we show that the frequency of the sing around sound soured is defined by the traveling time of the pulsed sound between a transmitter and a receiver. According to the motion of the sing around sound source, the traveling distance of the pulsed sound is increased. Therefore, the motion of the sing around sound source in air decreases the sing around frequency. The frequency modification according to the motion of the sound source depends on the relationship between the sound direction and the motion direction of the sound source. If the sing around sound source moves transverse to the sing around direction, i.e., $\theta = \pi/2$, where $\theta$ is the angle between the sound direction and the motion direction of the sound source shown in **Fig. 2**, we obtain $(c_A t_1)^2 = L^2 + (v t_1)^2$.

Thus,

$$t_1 = \frac{L}{\sqrt{c_A^2 - v^2}} = \frac{L/c_A}{\sqrt{1-\left(\frac{v}{c_A}\right)^2}} = \frac{t_0}{\sqrt{1-\left(\frac{v}{c_A}\right)^2}} \quad (1),$$

where $c_A$ is the sound speed, $t_1$ is the time when a pulsed sound arrives at position R, L is the path between a transmitter and a receiver when the sing around sound source is at rest, and v is the velocity of the sing around sound source, $t_0 = L/c_A$ is the reference time at rest, whereas $t_1$ is that in motion. The representation of the modified frequency is similar to that of the Lorentz transformation.

Fig. 2  Sound path of the sing around sound source in motion. The direction of motion is transverse to the direction of sound, i.e., $\theta = \pi/2$. Sing around frequency can be determined. It is because the going and returning paths are equal.

If the sing around sound source moves parallel to the sound direction, the sing around frequency cannot be determined, because the going and returning traveling times of the sound are different. **Figure 3** shows that at θ = 0, the going time $t_2$ is $t_2 = \dfrac{L}{c_A - v}$ and the returning time $t_3$ is $t_3 = \dfrac{L}{c_A + v}$; thus, $t_2$ is not equal to $t_3$. Therefore, we cannot define the sing around frequency. The sing around frequency has a resonance whose sharpness depends on the velocity v and angle θ. The resonance condition is θ = π/2, and the larger the velocity v, the sharper the resonance.

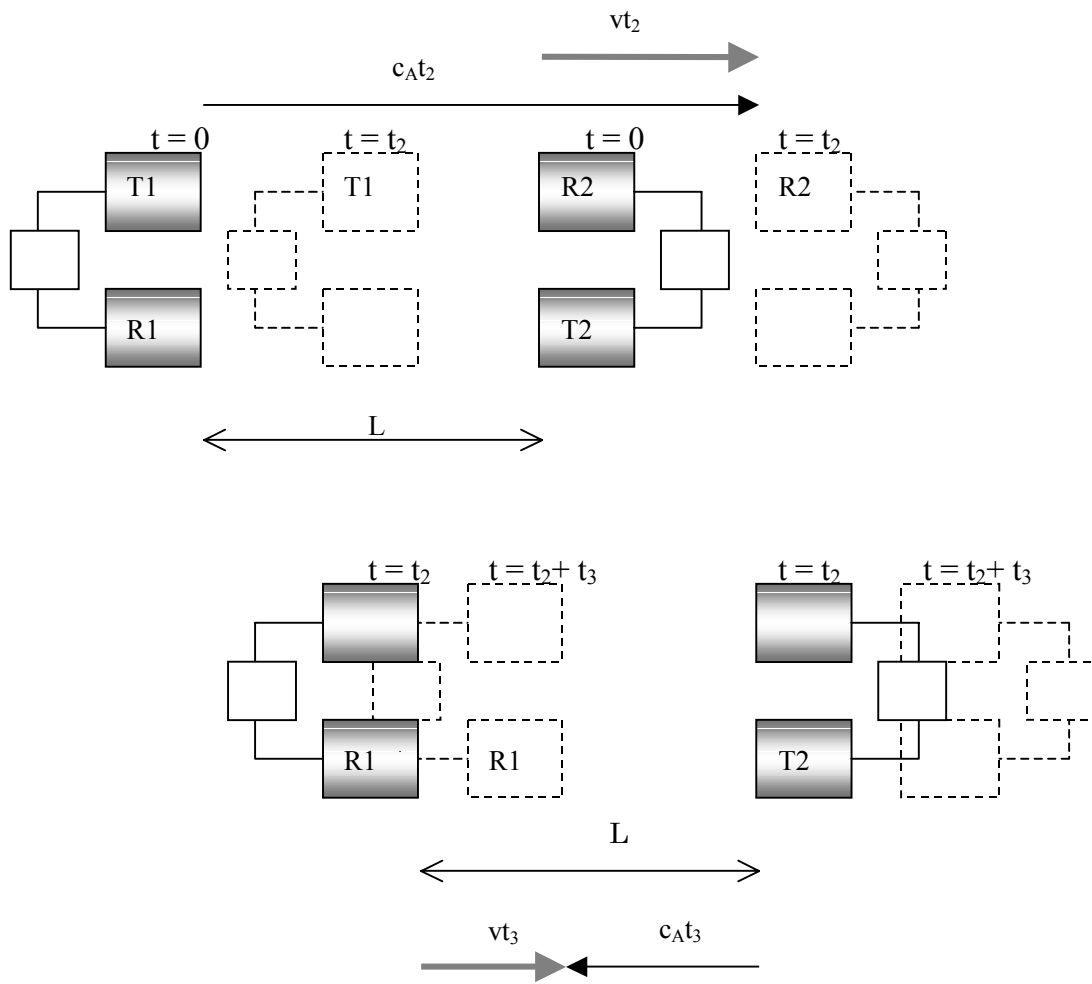

Fig. 3   Sound path of sing around sound source in motion. The direction of motion is parallel to the direction of sound. Going time $t_2$ is the period that pulsed sound transmitted from T1 is received by R2, going path is $c_A t_2$. We obtain the equation $c_A t_2 = L + v t_2$. Returning time $t_3$ is the period that pulsed sound transmitted from T2 (at the time $t_2$) is received by R2 (at the time $t = t_2 + t_3$). Returning path is $c_A t_3$, we obtain the equation $c_A t_3 = L - v t_3$, $t_3$ is not equal to $t_2$.

As mentioned above, the sing around frequency is defined as the frequency of repetitions. The frequency of repetitions is used as a fundamental (reference) frequency which can be used to define "time". We note here the analogy between the sing around sound source and the atomic clock, in that the reference frequency is determined on the basis of the velocity of the motion v as the representation of the Lorentz transformation.

We propose that the reference time, which is defined by the fundamental frequency, of the atomic clock is similar to the frequency definition mechanism of the sing around sound source.

B. Atomic clock in motion

There are similarities between the sing around sound source and the atomic clock. We can apply the concept of the sing around sound source to the atomic clock. The reference time of the atomic clock in motion can be determined on the basis of the flight time of photons. The photon in the atom in motion travels a longer distance than the atom at rest. Let us consider the electromagnetic phenomenon in an atom. This phenomenon occurs through a photon.

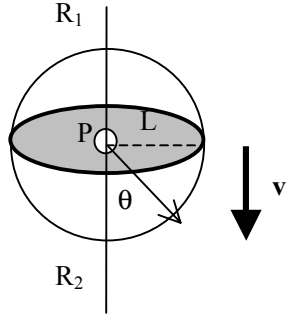

Fig. 4   Sound direction and the direction of source motion. Acoustic waves are irradiated from point sound source P for all directions. The intensity of sing around frequency is proportional to $\sin\theta$, for example, at $\theta = \pi/2$ reflection points are on the circle $2\pi L$, at $\theta = 0$ reflection points are only $R_1$ and $R_2$. Resonance condition is $\theta = \pi/2$, i.e., on this condition going time and returning time is equal, then sing around frequencies can be defined, whose frequency is represented by Lorentz transformation. According to the analogy of sing around sound source, we can define the time of atomic clock in motion.

Let us discuss the flight time of a photon in an atomic clock. As mentioned above, we assume that information is transmitted through photon; the time of the atomic clock is then defined by the flight time of the photon. Photons in an atomic clock in motion have to travel a longer distance than those in an atomic clock at rest. If the atomic clock moves at the velocity v, the photons travel different distances according to the direction of flight. Photons in an atomic clock at rest are emitted in all directions and have the same traveling distance independent of their directions. However, in motion, the travel distance of the photons depends on their direction. **Figure 4** shows the probability of photon arrival, where θ is the direction between the atomic clock motion and the traveling photon. The probability of photon arrival depends on sinθ. At θ =π/2, as mentioned in section 2.A, we obtain $(ct_1)^2 = L^2 + (vt_1)^2$; thus, equation (2), which is similar to equation (1), is

$$t_1 = \frac{L}{\sqrt{c^2 - v^2}} = \frac{L/c}{\sqrt{1-\left(\frac{v}{c}\right)^2}} = \frac{t_0}{\sqrt{1-\left(\frac{v}{c}\right)^2}} \quad (2),$$

where c is light the speed, $t_1$ is the time when a photon arrives at position R, L is the photon path in the atomic clock at rest, and v is the velocity of the atomic clock in motion, $t_0$ = L/c is the reference time in the atomic clock at rest, whereas $t_1$ is the reference time in the atomic clock in motion.

The probability of photon arrival is proportional to sinθ, and the most frequent case is at θ = π/2. This indicates that the probability that a photon moves parallel or antiparallel to the atom motion is very low. At θ = π/2, we obtain a high probability.

3. DISCUSSION

We discuss electromagnetic phenomena, for example, the radiation of an atomic clock. Electromagnetic phenomena occur through photons, i.e., they are exchanges of photons. According to equation (2), the photon path L is geometrically expanded when the atomic clock is in motion. (Here, L is for an atom radius.) Therefore, it takes more time in the atomic clock in motion than in the atomic clock at rest. We assume that the exchanges of photons occur resonantly. It is because only the photon moving transversely to the atom motion direction can be modified as shown in equation (2). The photon moving parallel to the atom motion direction cannot participate in electromagnetic phenomena, because the resonant frequency cannot be determined. Of course the resonant exchange of photons is a highly speculative hypothesis, however large amount of experimental data show that equation (2) is entirely accurate, indicating that these data are in accordance with the hypothesis.

The representation of "photon exchange" as well as that of "resonance", is ambiguous. Testing the hypothesis by experiment is not feasible; however, simple geometrical representation is attractive. We can directly derive the Lorentz transformation equation for time delay geometrically, which is different from the conventional technique. The conventional technique of deriving the Lorentz transformation equation is shown in the appendix.

4. CONCLUSION

In this letter, we showed that time is a combination of light speed and motion. We discussed the time in the atomic clock in motion, using the analogy of the sing around sound source. The sing around frequency is modified according to the motion of the sing around sound source, using the Lorentz transformation equation. If we use the sing around frequency as the reference, we can define the reference "time". We propose that the time delay of the atomic clock in motion be derived using the sing around method.

**References**

1) R. P. Feynman, R. B. Leighton, and M. Sands, *The Feynman Lectures on Physics* Vol. 3 (Addison Wesley, Reading, MA, 1965).
2) M. Sato, "Doppler shift of acoustic waves and Lorentz transformation," IEICE, [Ultrasonic], US2004-9, (in Japanese).
3) M. Sato, "Proposed experiment of local momentum transfer in Young's double slit," quant-ph/0406002.
4) M. Sato, "Proposed experiment on the continuity of quantum entanglement," quant-ph/0405155.
5) M. Sato, "Proposal of Signaling by Interference Control of Delayed-Choice Experimental Setup," quant-ph/0409059.

**Appendix**: Derivation of Lorentz transformation equation

From the results of the Michelson-Morley experiment, $2t_1 = t_2 + t_3$ was confirmed, where $t_1$ and $t_2$ are represented by equations (1) and (2) respectively. Therefore, the distance L is assumed to shrink. L is the distance measured vertically to the motion direction, and $L_L$ is the distance measured parallel to the motion direction,

$$\frac{2L}{\sqrt{c^2 - v^2}} = \frac{L_L}{c-v} + \frac{L_L}{c+v},$$

$$\therefore L_L = L\sqrt{1 - \left(\frac{v}{c}\right)^2}. \qquad (A.1)$$

This is the Lorentz transformation equation for distance.